\documentclass[journal]{IEEEtran}
\usepackage{amsmath}
\usepackage{amsfonts}
\usepackage{algorithmic}
\usepackage{algorithm}
\usepackage{array}
\usepackage[caption=false,font=normalsize,labelfont=sf,textfont=sf]{subfig}
\usepackage{xcolor}
\usepackage{textcomp}
\usepackage{stfloats}
\usepackage{url}
\usepackage{verbatim}
\usepackage{graphicx}
\usepackage{cite}
\usepackage{multirow}
\usepackage{booktabs}
\usepackage{comment}
\usepackage{color}
\begin{document}

\title{Towards Generalizable Deepfake Detection via Forgery-aware Audio-Visual Adaptation: A Variational Bayesian Approach}

\author{Fan Nie, Jiangqun Ni, ~\IEEEmembership{Member, ~IEEE}, Jian Zhang, Bin Zhang, Weizhe Zhang, ~\IEEEmembership{Senior Member, ~IEEE}, Bin Li, ~\IEEEmembership{Senior Member, ~IEEE}
        % <-this % stops a space
% \thanks{This paper was produced by the IEEE Publication Technology Group. They are in Piscataway, NJ.}% <-this % stops a space
% \thanks{Manuscript received April 19, 2021; revised August 16, 2021.}

%\thanks{This work was supported in part by the National Natural Science Foundation of China under Grants U23B2022 and U22A2030; in part by the Guangdong Major Project of Basic and Applied Basic Research under Grant 2023B0303000010; in part by the Major Key Project of PCL under Grant PCL2023A05. (\emph{Corresponding author: Jiangqun Ni.})}

\thanks{Fan Nie is with the School of Computer Science and Engineering, Sun Yat-sen University, Guangzhou 510006, China, and also with the Department of New Networks, Pengcheng Laboratory, Shenzhen 518066, China.}
\thanks{Jiangqun Ni is with the School of Cyber Science and Technology, Sun Yat-sen University, Shenzhen 510275, China, and also with the Department of New Networks, Pengcheng Laboratory, Shenzhen 518066, China (e-mail: issjqni@mail.sysu.edu.cn).}
\thanks{Jian Zhang is with the School of Computer Science and Cyber Engineering, Guangzhou University, Guangzhou 510006, China.}
\thanks{Bin Zhang is with the Department of New Networks, Pengcheng Laboratory, Shenzhen 518066, China.}
\thanks{Weizhe Zhang is with the School of Cyberspace Science, Harbin Institute of Technology, Harbin 150001, China, and also with the Department of New Networks, Peng Cheng Laboratory, Shenzhen 518066, China.}
\thanks{Bin Li is with the College of Information Engineering, Guangdong Key Laboratory of Intelligent Information Processing, and Shenzhen Key Laboratory of Media Security, Shenzhen University, Shenzhen 518060, China.}
}

% The paper headers
%\markboth{Journal of \LaTeX\ Class Files,~Vol.~14, No.~8, August~2021}%
\markboth{Submitted Manuscript for IEEE Transactions on Information Forensics and Security}%
{Shell \MakeLowercase{\textit{et al.}}: A Sample Article Using IEEEtran.cls for IEEE Journals}

% \IEEEpubid{0000--0000/00\$00.00~\copyright~2021 IEEE}
% Remember, if you use this you must call \IEEEpubidadjcol in the second
% column for its text to clear the IEEEpubid mark.

\maketitle

\begin{abstract}
The widespread application of AIGC contents has brought not only unprecedented opportunities, but also potential security concerns, e.g., audio-visual deepfakes. Therefore, it is of great importance to develop an effective and generalizable method for multi-modal deepfake detection. Typically, the audio-visual correlation learning could expose subtle cross-modal inconsistencies, e.g., audio-visual misalignment, which serve as crucial clues in deepfake detection. In this paper, we reformulate the correlation learning with variational Bayesian estimation, where audio-visual correlation is approximated as a Gaussian distributed latent variable, and thus develop a novel framework for deepfake detection, i.e., Forgery-aware Audio-Visual Adaptation with Variational Bayes (FoVB). Specifically, given the prior knowledge of pre-trained backbones, we adopt two core designs to estimate audio-visual correlations effectively. First, we exploit various difference convolutions and a high-pass filter to discern local and global forgery traces from both modalities. Second, with the extracted forgery-aware features, we estimate the latent Gaussian variable of audio-visual correlation via variational Bayes. Then, we factorize the variable into modality-specific and correlation-specific ones with orthogonality constraint, allowing them to better learn intra-modal and cross-modal forgery traces with less entanglement. Extensive experiments demonstrate that our FoVB outperforms other state-of-the-art methods in various benchmarks.
\end{abstract}

\section{Introduction}

Nowadays, advanced AI generative content (AIGC) techniques enable the creation of diverse and realistic multimedia content. Given the corresponding audio, for example, it could incorporate the arbitrary image to create an audio-visually convincing video with near-perfect audio-visual synchronization. Leveraging such techniques, malicious attackers may generate or manipulate sensitive contents, e.g., synthetic facial images (deepfake), to spread harmful misinformation, thusdamaging individual reputations and cheating in cyberspace. Therefore, developing an effective detection method for audio-visual deepfakes is a longstanding goal for the forensic community.

\begin{figure}[t]
    \centering
    \includegraphics[width=\linewidth]{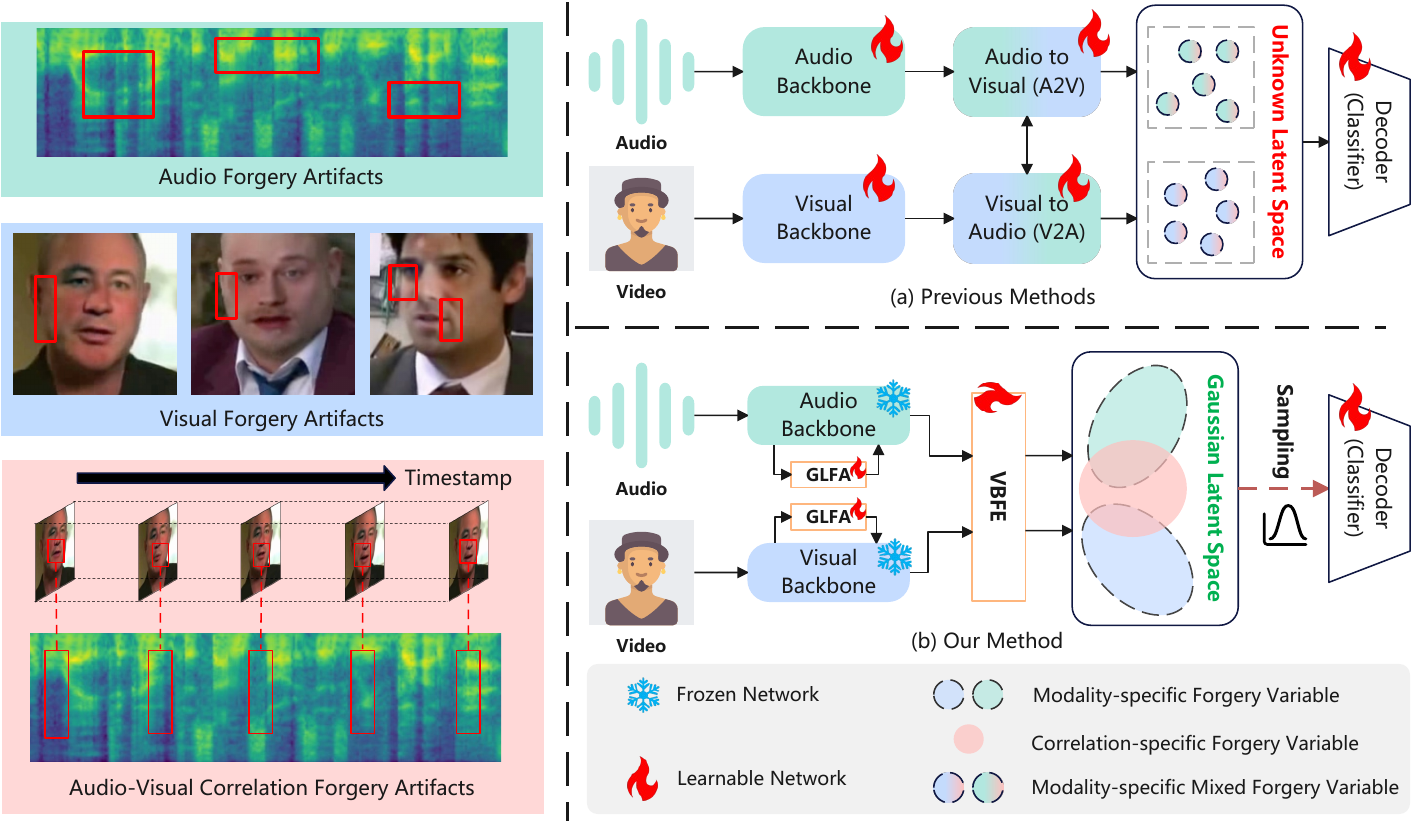}
    \caption{\textbf{Difference between the proposed FoVB and previous methods.} Previous methods leverage massive samples to fully fine-tune the pre-trained backbones, e.g., ViT, ResNet, etc, and capture forgery artifacts, whose latent variables are distributed as unknown. In contrast, with the frozen pre-trained backbones, our FoVB adopts two core designs, e.g., Glocal-Local Forgery-aware Adaptation (GLFA) and Variational Bayesian Forgery Estimation (VBFE), to capture intra-modal (audio or visual) and cross-modal (audio-visual) forgery traces characterized by Gaussian distributed latent variables and thus learn more generalizable forgery representation.}
    \label{fig:main_difference}
\end{figure}

Most existing methods \cite{zheng_exploring_2021, shiohara_detecting_2022, mullerDoesAudioDeepfake2022} detect subtle deepfake traces from either the audio or visual modality, which prevents them from meeting the practical demands of audio-visual forensics. In general, as shown in Fig.\ref{fig:main_difference}, besides uni-modal forgeries, i.e., spectrum abnormalities of audio signal, facial boundary artifacts, and illumination inconsistency on facial regions, the audio-visual deepfake generation will inevitably introduce some forgery artifacts in terms of audio-visual inconsistencies, e.g., imperfect lip movement \cite{haliassos_lips_2021, prajwal_lip_2020}, phoneme-viseme mismatch \cite{Agarwal_2020_CVPR_Workshops}, and emotion dissimilarity \cite{mittal_emotions_2020, bai_aunet_2023}. Accordingly, several multi-modal design methods are presented in \cite{zhouJointAudioVisualDeepfake2021, Agarwal_2020_CVPR_Workshops, cai_you_2022, yu_pvass-mdd_2023, yang_avoid-df_2023, liu_mcl_2023} to capture audio-visual inconsistencies. While in \cite{yu_pvass-mdd_2023, yang_avoid-df_2023, oorloff_avff_2024}, the pre-trained backbones, e.g., Vision-Transformer (ViT), are fine-tuned with a small amount of audio-visual forgery samples to learn the more generalizable forgery representation.

Despite their promising representation capabilities, existing transformer-based detectors still fall short in the following aspects: 
(1) These methods heavily depend on the fine-tuning of the full parameters of the pre-trained ViTs with audio-visual samples. Considering that the sizes of audio-visual datasets are generally much smaller than those for ViT pre-training, the detectors may overfit to specific artifacts and lose the ability to recall the generalized knowledge learned from pre-training, thus preventing them from generalizing well to unseen forgeries.  
(2) Existing methods leverage a large amount of audio-visual training pairs to learn discrete latent representations of audio-visual correlation for discriminating multi-modal deepfake detection. Nevertheless, the learned discrete representations of audio-visual correlations may not well characterize the complex audio-visual correlations, which prevents detectors from capturing unseen cross-modal inconsistencies.

Based on the above observations, we propose a multi-modal framework, i.e., Forgery-aware Audio-Visual Adaptation with Variational Bayes (FoVB), to detect audio-visual deepfakes. Specifically, as shown in Fig.\ref{fig:main_difference}, to preserve the built-in general knowledge of pre-trained backbones and effectively learn forgery-relevant features, we freeze the network parameters of the pre-trained backbones while introducing the following adaptation designs: (1) \textit{Global-Local Forgery-aware Adaptation.} Recognizing the fact \cite{qian_thinking_2020, yang_avoid-df_2023} that, for fake audio-visual contents, their subtle artifacts usually appear in the textural regions, i.e., the high-frequency components of visual content and audio spectrograms, we thus incorporate various difference convolutions and the a high-pass filter to capture intra-modal forgery features from both global and local perspectives. (2) \textit{Variational Bayesian Forgery Estimation.}Unlike previous arts, we enforce a Gaussian distribution on the latent variable to approximate the audio-visual correlations and learn the distribution with variational Bayesian estimation. To take advantage of the audio-visual forgery traces with variational Bayes, we factorize the Gaussian distributed latent variable into modality-specific and correlation-specific ones with an orthogonal constraint, allowing them to learn intra-modal and cross-modal forgery traces with less entanglement, which is beneficial for improving model generalization \cite{sutter_multimodal_2020, mao_multimodal_2023, nie2024frade}. The estimated latent variables are then sampled and adapted into the pre-trained backbones to facilitate the learning of generalizable forgery representations.

The main contributions of the paper are summarized as follows:
\begin{itemize}
    \item A novel audio-visual deepfake detection framework, i.e., Forgery-aware Audio-Visual Adaptation with Variational Bayes (FoVB), is proposed to adapt forgery-relevant features into the pre-trained backbones.
    \item We exploit various forgery-aware difference convolutions and a high-pass filter to extract local and global forgery traces in the audio and visual modalities. With the extracted features, we enforce a Gaussian-distributed latent variable of audio-visual traces, which is further decomposed into fine-grained modality-specific and correlation-specific variables with orthogonal constraint, and estimated with variational Bayes.
    \item Extensive experiments demonstrate that the proposed FoVB outperforms the existing counterparts in terms of generalizability and robustness performance.
\end{itemize}

\section{Related Work}

\subsection{Deepfake Detection} Nowadays, generic uni-modal detection methods can not meet the forensic demands of advanced AIGC techniques, which simultaneously manipulate audio and visual content to deliver synthetic media. Therefore, the audio-visual inconsistencies are exploited to learn more discriminative representations for multi-modal forensics. Recent works \cite{chugh_not_2020, yang_avoid-df_2023, liu_mcl_2023, yu_pvass-mdd_2023} take advantages of joint learning framework to learn the generalized and intrinsic discrepancies between the two modalities. In \cite{feng_self-supervised_2023, oorloff_avff_2024}, the detectors are pre-trained in a self-supervised manner with large-scale human talking datasets, e.g., LRS2 \cite{afourasDeepAudiovisualSpeech2022} and LRS3 \cite{afourasLRS3TEDLargescaleDataset2018}, to learn the latent representation of audio-visual correlation. However, in the above self-supervised paradigm, the estimation of latent correlation heavily relies on the amount of audio-visual samples, and could be considered discrete and not well regularized, thus limiting the detectors' generalizability. Different from the self-supervised methods, our FoVB assumes the latent variables of audio-visual deepfakes to be Gaussian distributed and drives the optimization objectives within variational Bayesian framework.

\subsection{Parameter-efficient Audio-Visual Representation Learning}
Parameter-efficient fine-tuning \cite{houlsby_parameter-efficient_2019} (PEFT) is the method to adapt the pre-trained backbones to downstream tasks with better generalizability by updating a small subset of the model's parameters while keeping the majority of the pre-trained weights frozen. Following the paradigm of PEFT, there are a range of methods \cite{pan_st-adapter_2022, lin_vision_2023, duan_cross-modal_2023, Wang_2024_CVPR, Liang_2024_CVPR} which exploit pre-trained transformer backbones and learnable parameters, including convolution adapter \cite{pan_st-adapter_2022}, cross-modal adapter \cite{lin_vision_2023, duan_cross-modal_2023}, learnable prompting \cite{Liang_2024_CVPR}, to extract audio-visual representations for downstream tasks and achieve promising results. In \cite{shao_deepfake-adapter_2023, luo_forgery-aware_2023, kong_moe-ffd_2024}, the locality-aware adapters are adopted to characterize forgery artifacts for uni-modal deepfake detection. For audio-visual deepfake detection, the cross-modal inconsistencies between audio-visual modalities should be fully explored. Therefore, the FoVB framework is proposed which incorporates various forgery-aware convolutions to expose intra-modal forgery traces and adopts variational Bayes to characterize the audio-visual correlations for generalizable multi-modal deepfake detection.

\subsection{Multi-modal Variational Bayes}
A prime example of uni-modal variational Bayes is the Variational autoencoder (VAE) \cite{kingma_auto-encoding_2014, sohn_learning_2015}, a generative model featured by the latent variable for single modality and could be obtained by maximizing the evidence lower bound (ELBO). In general, the ELBO can be decomposed into a reconstruction term and a KL-divergence term, which measures the difference between the variational posterior and the prior distribution of the latent variable. While for multi-modal variational Bayes, e.g., multi-modal VAE (MVAEs), the estimation of KL-divergence between the joint variational posterior and the joint prior distribution of all involved modalities is generally computationally intractable and can be approximated by the mixture of experts (MoE) \cite{shi_variational_2019, mao_multimodal_2023}. The MVAE has found applications in various fields, e.g., audio-visual segmentation \cite{mao_multimodal_2023}, fake news detection \cite{multi_vae_fake_news}, and emotion detection \cite{multi_vae_EGG} by taking advantage of the shared latent representation among different modalities. Our method, however, concentrates on characterizing forgery-relevant latent representations from both modality-specific and correlation-specific perspectives, and imposes an orthogonality constraint to facilitate the learning of both intra-modal and cross-modal forgery traces.

\section{Method}

\begin{figure*}[ht]
    \centering
    \includegraphics[width=0.8\textwidth]{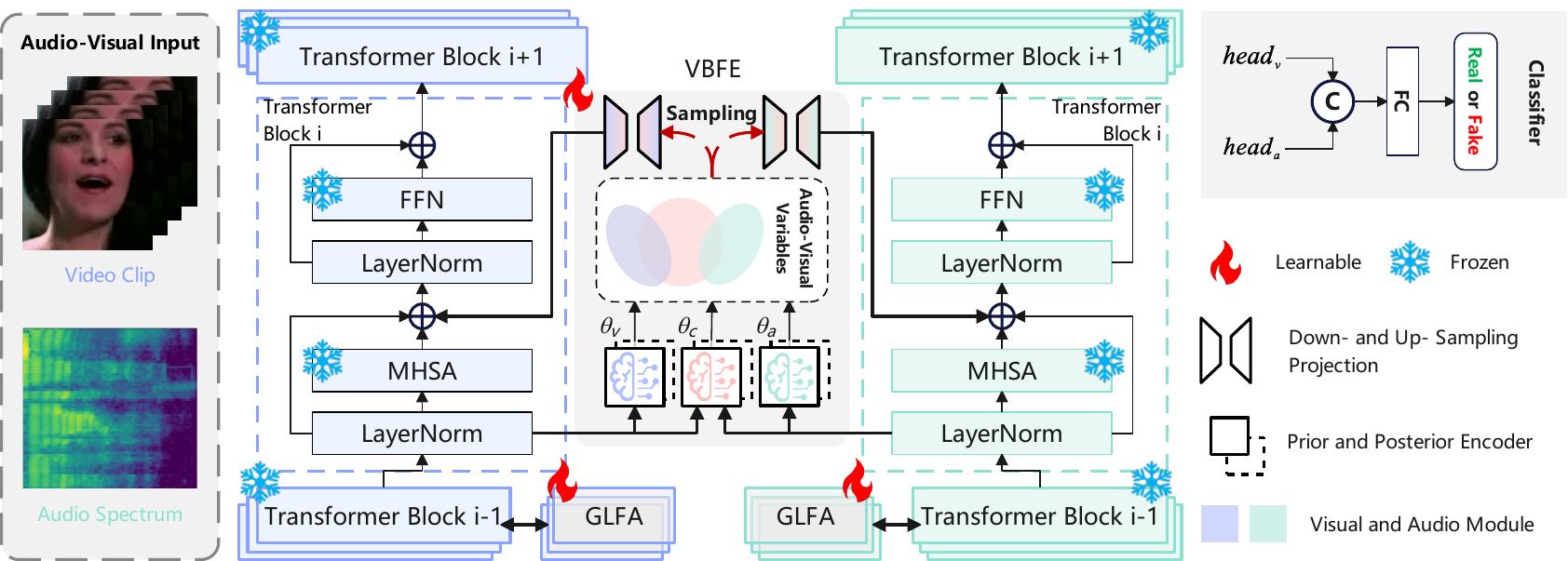}
    \caption{\textbf{Overview of the proposed FoVB framework.} The audio-visual sequences are fed into pre-trained backbones, where the GLFA extracts intra-modal forgery features, and the VBFE exploits the extracted features by the $i$-th transformer block to estimate audio-visual latent variables with variational Bayes. Then, the estimated variables facilitate the adaptation of forgery-relevant knowledge. Finally, we leverage the classification heads of audio-visual sequences, i.e., $\mathrm{head}_{v}$ and $\mathrm{head}_{a}$, to determine if forgery exists.}
    \label{fig:overview}
\end{figure*}

This section presents our proposed Forgery-aware Audio-Visual Adaptation with Variational Bayes (FoVB) framework that adapts pre-trained ViTs to detect audio-visual deepfakes with additional learnable parameters. Concretely, as illustrated in Fig.\ref{fig:overview}, we design two adapter modules, i.e., Global-Local Forgery-aware Adaptation (GLFA) and Variational Bayesian Forgery Estimation (VBFE) modules. The former is deployed into shallow transformer blocks to extract intra-modal forgery features, while the latter exploits extracted features to estimate the audio-visual variables in latent space.

\subsection{Audio-Visual Input Tokenization}
We consider a video clip $V \in \mathbb{R}^{T \times H_{v} \times W_{v} \times 3}$, consisting of $T$ RGB frames with resolution $H_{v} \times W_{v}$. The associated audio signal is transformed into 2D spectrogram $A \in \mathbb{R}^{H_{a} \times W_{a}}$, which is then replicated to $\hat{A} \in \mathbb{R}^{H_{a} \times W_{a} \times 3}$, similar to \cite{feng_self-supervised_2023, yang_avoid-df_2023, oorloff_avff_2024}. Given the $P \times P$ patch size and embedding projection for the adapted transformer backbone, we patchify and embed audio-visual inputs into two $D$-dim sequences with classification heads, i.e., $x_{v} \in \mathbb{R}^{(N_{v}+1) \times D}, x_{a} \in \mathbb{R}^{(N_{a}+1) \times D}$, where $N_{a}$ and $N_{v}$ denote the number of visual and audio patches, respectively. Note that, the classification tokens are not involved in the forward propagation of the following GLFA and VBFE modules.

\subsection{Global-Local Forgery-aware Adaptation}

To learn forgery-relevant features with pre-trained backbones, e.g., ViT, we exploit the audio-visual forgery artifacts that commonly appear in the high-frequency components of the local textural regions by the Global-Local Forgery-aware Adaptation (GLFA) module. In specific, diverse difference convolutions and high-pass filters are employed to expose intra-modal forgery traces from both the global and local perspectives. These forgery-aware features are then transformed with learnable projections into offsets of the original query$(q)$, key$(k)$, and value$(v)$ embeddings, i.e., $\Delta q, \Delta k,$ and $\Delta v$, and integrated with $q, k,$ and $v$ via element-wise addition to obtain $q+\Delta q, k+\Delta k, v+\Delta v$ before Multi-Head Self-Attention (MHSA). Compared with the Feed-Forward Network (FFN) in the ViT, MHSA exhibits a better representation capability, which is more suitable for learning of forgery traces with PEFT.

Initially, the input audio (visual) sequence $x \in \mathbb{R}^{N \times D}$ is reshaped into the spatial form $\hat{x} \in \mathbb{R}^{h \times w \times D}$ with $(h,w) = (\frac{H}{P}, \frac{W}{P})$, where $(H, W) = (H_{a}, W_{a})$ for the audio sequence $x_{a}$. The reshaped $\hat{x}$ is down-sampled with learnable projection into $\hat{x}_{down} \in \mathbb{R}^{h \times w \times \frac{D}{r}}$, where $r$ is the down-sampling factor. We then employ several well-designed convolutions \cite{liuExtendedLocalBinary2012, liuFusingSortedRandom2015, suBIRDLearningBinary2019}, including Angular Difference Convolution (ADC), Central Difference Convolution (CDC), Radial Difference Convolution (RDC), and Second-Order Convolution (SOC) \cite{fei_learning_2022}, as depicted in Fig.\ref{fig: GLFA_kernels}, to extract local textural features. Meanwhile, we apply the global high-frequency convolution (GFC) \cite{qian_thinking_2020}, which incorporates the Fast Fourier Transformation (FFT) and its inverse (IFFT) with a high-pass mask $M_{f}$ to discard low-frequency components and enforce the detector to concentrate on the high-frequency forgery artifacts. The extracted global and local features are then concatenated and converted into the forgery-aware offsets $\Delta q, \Delta k$, and $\Delta v$ via linear projections. Accordingly, the implementation of GLFA is described in Alg. \ref{alg:glfa}.

\begin{algorithm}
\caption{GLFA Implementation}
\label{alg:glfa}
\renewcommand{\algorithmicrequire}{\textbf{Input:}}
\renewcommand{\algorithmicensure}{\textbf{Args:}}
\renewcommand{\algorithmiccomment}[1]{\hfill $\triangleright$ #1}
\begin{algorithmic}
\REQUIRE modality sequence $x$, query embedding $q$, key embedding $k$, value embedding $v$.
\ENSURE convolution sets {\footnotesize $OpSet \in \{ADC, CDC, RDC, SOC\}$}, Fourier Transformation $FFT$, inverse transformation $IFFT$, high-pass mask $M_{f}$, linear projections $T_{q}, T_{k}, T_{v}$, multi-head self-attention $MHSA$.
\STATE $\hat{x} \gets Resize(x)$ \COMMENT{$\hat{x} \in \mathbb{R}^{h \times w \times D}$}
\STATE $\hat{x}_{down} \gets DownSample(\hat{x}, r)$ \COMMENT{$\hat{x}_{down} \in \mathbb{R}^{h \times w \times \frac{D}{r}}$} 
\FOR{$Op$ in $OpSet$}
\STATE $\hat{x}_{Op} \gets Op(\hat{x}_{down})$ 
\ENDFOR
\STATE $\hat{x}_{GFC} \gets IFFT(M_{f} \ast FFT(\hat{x}_{down}))$ \COMMENT{$\ast$: element-wise multiplication} 
\STATE $\hat{x}_{c} \gets Concat(\hat{x}_{ADC}, \hat{x}_{CDC}, \hat{x}_{RDC}, \hat{x}_{SOC}, \hat{x}_{GFC})$.
\STATE $\Delta q \gets T_{q}(\hat{x}_{c})$
\STATE $\Delta k \gets T_{k}(\hat{x}_{c})$
\STATE $\Delta v \gets T_{v}(\hat{x}_{c})$
\STATE $x_{out} \gets MHSA(q+\Delta q, k+\Delta k, v+ \Delta v)$
\RETURN $x_{out}$

\end{algorithmic}
\end{algorithm}

\begin{comment}
{\footnotesize
\begin{flalign}
    &\ \hat{x}_{f} = \mathrm{IFFT}(M_{f} \ast \mathrm{FFT}(\hat{x}_{down})), & \\
    &\ \hat{x}_{o} = o(\hat{x}_{down}), o \in \{\mathrm{ADC}, \mathrm{CDC}, \mathrm{RDC}, \mathrm{SOC}\}, & \\
    &\ x_{c} = \left[ \hat{x}_{f}, \hat{x}_{\mathrm{ADC}}, \hat{x}_{\mathrm{CDC}}, \hat{x}_{\mathrm{RDC}}, \hat{x}_{\mathrm{SOC}} \right], & \\
    &\ \delta q, \delta k, \delta v = \mathrm{conv}_{1 \times 1}(x_{c}), \mathrm{conv}_{1 \times 1}(x_{c}), \mathrm{conv}_{1 \times 1}(x_{c})& \\
    &\ \hat{x}_{out} = \mathrm{MHSA}(q+ \delta q, k + \delta k, v + \delta v), &
\end{flalign}}
where $\delta q, \delta k,$ and $\delta v$ represent the forgery-aware offsets.
\end{comment}

 Since ViT backbones commonly perform multi-head projection when generating the query, key, and value embeddings in MHSA, we project the intermediate features $x_{c}$ into offset embeddings in the multi-head form. 

\subsection{Variational Bayesian Forgery Estimation\protect\footnote{See the Appendix for complete derivation.}}
\subsubsection{Preliminaries}
\paragraph{Variational Inference} Variational (Bayesian) Inference can be used to infer the posterior distribution over latent variables given the observations (and parameters). With the input $X$ and the label $y$, the generative process is to draw the latent variable $z$ from the prior distribution $p_{\theta}(z\vert X)$, which is parameterized with $\theta$, and generate the output via $p_{\theta}(y \vert X,z)$. The inference process could be formulated as the estimation of the distribution for the latent variable $z$, via the observed $X, y$ by computing the posterior $p_{\theta}(z \vert X, y)$, which is computationally intractable and thus approximated with simplified distribution $q_{\phi}(z \vert X, y)$ parameterized by $\phi$. Therefore, the goal of inference is to find the optimal $\hat{\theta}$ and $\hat{\phi}$ to maximize the following log-likelihood objective:

{\footnotesize
\begin{equation}
    \begin{split}
         \{\hat{\theta}, \hat{\phi}\} &= \arg \underset{\theta, \phi}{max} \, \mathrm{log} \, {p_{\theta}(y \vert X)} \\
         &= \mathbb{E}_{q_{\phi}(z \vert X,y)} \, \mathrm{log} \, p_{\theta}(y \vert X, z) - D_{KL}(q_{\phi}(z \vert X, y) \Vert p_{\theta}(z \vert X)) \\
         & + D_{KL}(q_{\phi}(z \vert X, y) \Vert p_{\theta}(z \vert X, y)), \\
    \end{split}
     \label{equ:raw_optim}
\end{equation}
}

\noindent where the log-likelihood term is converted with the approximated posterior $q_{\phi}(z \vert X, y)$ and the latent variable $z$. Notably, the KL divergence term ($D_{KL}$) is always greater than or equal to zero, and thus maximizing $\mathrm{log} \, p_{\theta}(y \vert X)$ could be achieved by maximizing the evidence lower bound (ELBO):

{\footnotesize
\begin{equation}
    \begin{split}
         \{\hat{\theta}, \hat{\phi}\} &= \arg \underset{\theta, \phi}{max} \, \mathrm{log} \, {p_{\theta}(y \vert X)} \\
         &= \arg \underset{\theta, \phi}{max} \, \mathrm{ELBO}(X,y,\theta,\phi) \\
         &= \arg \underset{\theta, \phi}{max} \,\mathbb{E}_{q_{\phi}(z \vert X,y)} \, \mathrm{log} \, p_{\theta}(y \vert X, z) - D_{KL}(q_{\phi}(z \vert X, y) \Vert p_{\theta}(z \vert X)), \\
    \end{split}
     \label{equ:elbo_optim}
\end{equation}
}

\noindent where the KL term could be solved in closed form with the assumption that the prior $p_{\theta}(z \vert X)$ and posterior $q_{\phi}(z \vert X, y)$ are Gaussian distributed. 

\paragraph{Multi-modal Variational Inference} In the multi-modal setting, the latent variable $z$ is derived from $K$ modalities, i.e., $X = \{x_{i}\}_{i=1}^{K}$. To optimize the distribution of $z$, it could be approximated as the mixture of experts (MoE) by factorizing the joint prior (posterior) distribution as the weighted sum of uni-modal priors (posteriors) \cite{shi_variational_2019}. Accordingly, the $D_{KL}$ term in $\mathrm{ELBO}(X,y,\theta, \phi)$ is the lower bound of the weighted sum of independent KLs:

{\footnotesize
\begin{equation}
    \begin{aligned}
     D_{KL} (q_{\phi}(z \vert X,y) \Vert p_{\theta}(z \vert X)) \leq \\
     \sum_{i=1}^{K} \pi_{i} D_{KL}(q_{\phi_{i}}(z \vert x_{i}, y) \Vert p_{\theta}(z \vert x_{i})),
    \label{equ:low_bound_moe}
    \end{aligned}
\end{equation}
}

\noindent where $\sum_{i=1}^{K} \pi_{i} = 1$. However, the correlations among various modalities are not considered in the MoE setting of Eq. \ref{equ:low_bound_moe}, which is contrary to our intention of modeling the multi-modal correlation with the latent variable $z$. Therefore, we adopt the MoE-based dynamic prior $f_{K}$ \cite{sutter_multimodal_2020} to replace each uni-modal prior $p_{\theta}(z \vert x_{i})$ in Eq, \ref{equ:low_bound_moe} with the arithmetic mean of all uni-modal priors and posterior, i.e., 

{\footnotesize
    \begin{equation}
    f_{K} = \frac{1}{2} \left (\sum_{i=1}^{K} \pi_{i} q_{\phi}(z \vert x_{i}, y) + \sum_{i=1}^{K} \pi_{i} p_{\theta}(z \vert x_{i}) \right ).
    \end{equation}
}

The rationale behind the $f_{K}$ is summarized as follows. Compared with the uni-modal prior $p_{\theta}(z \vert x_{i})$, the dynamic prior incorporates the prior knowledge that there exists shared component among each modality, in other words, the multi-modal label $y$ is determined by the latent variable $z$, and allows $z$ to model the multi-modal correlation in the variational Bayesian framework. We therefore call $f_{K}$ prior due to its role in the ELBO formulation and optimization.

Accordingly, we rewrite the KL divergence term of Eq. \ref{equ:elbo_optim} with the adopted $f_{K}$, and then derive a new lower bound $\widetilde{\mathrm{{ELBO}}}(X,y,\theta,\phi)$ over $\mathrm{ELBO}(X,y,\theta, \phi)$ with combinatorial Jensen-Shannon ($D_{JS}$) divergence as follows, detailed in Eq. \ref{equ: supp_js_lowbound}.

{\footnotesize
\begin{flalign}
    \begin{aligned}
        &\ \mathrm{ELBO}(X,y,\theta,\phi) &\\
        &\ \geq \mathbb{E}_{q_{\phi}(z \vert X, y)} \mathrm{log} \, p_{\theta}(y \vert X, z) - D_{JS}(q_{\phi}(z \vert X,y), p_{\theta}(z \vert X)) & \\
        &\ = \widetilde{\mathrm{{ELBO}}}(X,y,\theta,\phi).
    \end{aligned}
    \label{equ: js_lowbound}
\end{flalign}
}

\subsubsection{Forgery-aware Latent Variable Factorization}
Given the well-derived $\widetilde{\mathrm{{ELBO}}}(X,y,\theta,\phi)$, we could directly leverage this optimization objective to model the audio-visual correlations. Note that the multi-modal deepfake techniques not only disrupt the correlation between audio and visual modalities but also leave intra-modal inconsistency traces, which are not well characterized in the above $\widetilde{\mathrm{{ELBO}}}(X,y,\theta,\phi)$. In other words, the estimated audio-visual variable is subject to the modality-specific information, as shown \cite{sutter_multimodal_2020, mao_multimodal_2023, nie2024frade}. We then factorize the audio-visual variable $z$ into three fine-grained components, i.e., the correlation-specific, audio-specific, and visual-specific ones, to comprehensively evaluate their contributions to deepfake detection. 

In specific, for the audio-visual pair $X = \{x_{a}, x_{v}\}$ processed by GLFA and its forgery annotation $y$, we factorize the latent variable $z$ into two types of variables, i.e., the modality-specific $s = \{s_{a}, s_{v}\}$ and correlation-specific $c$. While the former learns the reliable and stable representation of intra-modal forgery traces, the latter is exploited to estimate the audio-visual correlation in latent space. On the other hand, there exist three categories of forgeries for audio-visual deepfakes based on different combinations of audio and visual deepfakes, i.e., FakeVisual-FakeAudio (FVFA), FakeVisual-RealAudio (FVRA), and RealVisual-FakeAudio (RVFA). However, the aforementioned annotation does not really give the fine-grained descriptions for various forgery combinations, which may mislead the estimation of modality-specific variables and introduce estimation errors for audio-visual inconsistencies that are simply impossible to ignore. Therefore, we define the forgery annotations of individual audio and visual modalities as $y_{a}, y_{v}$, which are incorporated with the factorized variables $s$ and $c$, and rewrite the $\widetilde{\mathrm{{ELBO}}}(X,y,\theta,\phi)$ as follows:

{\footnotesize
\begin{flalign}
    \begin{aligned}
        &\ \widetilde{\mathrm{{ELBO}}}(X,y,\theta,\phi) & \\
        &\ = \sum_{o\in \{a,v\}} \mathbb{E}_{q_{\phi_{c}}(c \vert X, y)} \left [ \mathbb{E}_{q_{\phi_{s_{o}}}(s_{o} \vert x_{o}, y_{o})} \left [ \mathrm{log} \, p_{\theta}(y \vert x_{o}, s_{o}, c) \right] \right] & \\
        &\ - \sum_{o \in \{a,v\}} D_{KL}(q_{\phi_{s_{o}}}(s_{o} \vert x_{o},y_{o}) \Vert p_{\theta}(s_{o} \vert x_{o})) & \\
        &\ - D_{JS}(q_{\phi}(c \vert X,y), p_{\theta}(c \vert X)). &
    \end{aligned}
    \label{equ: factorized_js_elbo}
\end{flalign}
}

\subsubsection{Audio-Visual Variable Adpatation}
The estimated Gaussian latent variables are then exploited to adapt to the pre-trained backbones. As shown in Fig.\ref{fig:overview}, the variables are sampled as the latent codes and transformed into variational modality representations. In specific, given the sampled codes $c \sim q_{\phi}(c \vert x_{a}, x_{v}, y) \in \mathbb{R}^{D}$, $s_{a} \sim q_{\phi}(s_{a} \vert x_{a}, y_{a}) \in \mathbb{R}^{D}$, and $s_{v} \sim q_{\phi}(s_{v} \vert x_{v}, y_{v}) \in \mathbb{R}^{D}$, we concatenate the correlation-specific code $c$ with both audio- and visual-specific codes to obtain the fused representations for each modality via linear projections, i.e., $\{\hat{sc}_{a}, \hat{sc}_{v}\} \in \mathbb{R}^{D}$. The $\hat{sc}_{a}, \hat{sc}_{v}$ are then combined respectively with outputs of MHSA in the audio and visual transformer blocks via element-wise addition.

\subsection{Optimization Objectives}
Incorporating the derived $\widetilde{\mathrm{{ELBO}}}(X,y,\theta,\phi)$ in Eq. \ref{equ: factorized_js_elbo}, we enforce the audio-visual orthogonality constraint to ensure the modality-specific and correlation-specific variables could learn intra-modal and cross-modal forgery traces  with less entanglement. Specifically, given the latent variables $c, s_{a}, $ and $s_{v}$, the orthogonality constraint is formulated as follows:

{\footnotesize
\begin{equation}
    \mathcal{L}_{\mathrm{orth}} = {\parallel c^{T}s_{a} \parallel}_{F}^{2} + {\parallel c^{T}s_{v} \parallel}_{F}^{2} + {\parallel s_{a}^{T}s_{v} \parallel}_{F}^{2},
\end{equation}
}

\noindent where ${\parallel \cdot \parallel}^{2}_{F}$ denotes the squared Frobenius norm. On the other hand, with the output sequences $x_{a}$ and $x_{v}$, we exploit their head tokens with linear projections to obtain the final prediction $\hat{y}$, which is optimized via cross-entropy loss $\mathcal{L}_{ce}$. The overall optimization objectives could be defined as follows:

{\footnotesize
\begin{equation}
    \mathcal{L} = \mathcal{L}_{ce}(\hat{y}, y) - \widetilde{\mathrm{{ELBO}}}(X,y,\theta,\phi) + \alpha \mathcal{L}_{\mathrm{orth}}.
    \label{equ: total_loss}
\end{equation}
}

\subsection{Framework Training and Inference}
According to the principle of variational Bayes, we have the posterior and prior variable encoders $\phi, \theta$ to estimate audio-visual variables in Gaussian latent space. In the training phase, as shown in the reconstruction term in Eq. \ref{equ: factorized_js_elbo}, we exploit the posterior estimation encoded by $\phi$ for label reconstruction, i.e., forgery prediction. And The prior and posterior distributions of the latent variables are expected to be close by decreasing their KL-divergence. During inference, given the audio-visual input, the VBFE receives the processed features from transformer blocks with GLFA and estimates the Gaussian latent variables via the encoder $\theta$. The well-devised variables are sampled and fused with the input features, which are then fed into successive transformer blocks of the backbone. Finally, the head tokens of the audio-visual sequence are leveraged to predict its authenticity.

\section{Experiments and Results}
\subsection{Implementation Details}
\paragraph{Input Preprocess and Training Settings} Visual frames and the corresponding audio signals are extracted from the raw videos at a sampling rate of 5 fps and 16 kHz. Following \cite{feng_self-supervised_2023, oorloff_avff_2024}, facial regions are extracted from the sampled frames by MTCNN \cite{zhang_joint_2016}. The audio signal is transformed into the log-mel spectrogram via short-term Fourier Transform with 80 mel filter banks, a hop length of 160, and a window size of 320. For framework training, the AdamW optimizer with a learning rate $2e-6$ is utilized, and the hyperparameter $\alpha$ in Eq.\ref{equ: total_loss} is set to 0.1.

\begin{figure}[tb]
    \centering
    \includegraphics[width=\linewidth]{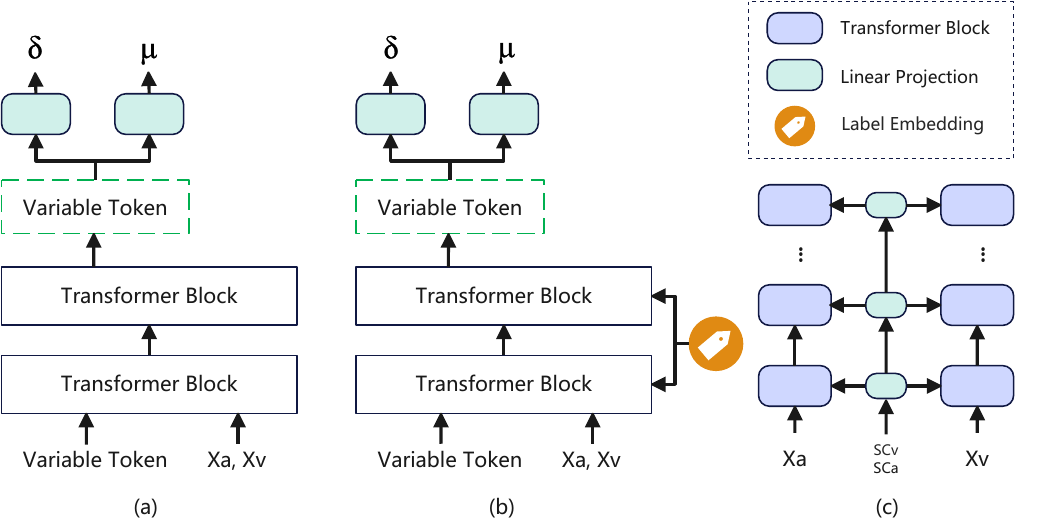}
    \caption{\textbf{Implementation of prior and posterior encoders in VBFE.} $x_{a}, x_{v}$ are fed into the encoders $\theta, \phi$ to estimate the mean and variance values $\mu, \delta$ of latent variables. (a) Architecture of the prior encoder $\theta$. (b) Architecture of the posterior encoder $\phi$, which incorporates label embedding with cross-attention operations to estimate the posterior distribution. (c) Pipeline of variable adaptation.}
    \label{fig: FoVB_impl}
\end{figure}

\paragraph{Implementation of the VBFE module} The prior encoder $\theta$ is designed to estimate latent variables based on audio-visual inputs, while the posterior encoder $\phi$ approximates latent variables using both audio-visual inputs and their associated labels. Specifically, consistent with the FoVB backbones, we utilize stacking identical transformer blocks to construct the aforementioned encoders. As illustrated in Fig.\ref{fig: FoVB_impl}, both variable encoders exploit an extra $D$-dim variable token and linear projections to derive the Gaussian mean and variance of the variables. For Audio-Visual Variable Adaptation, the $sc_{a}, sc_{v}$ are sampled from the estimated Gaussian variables, and fused into the outputs $x_{a}, x_{v}$ of subsequent transformer blocks via linear projection and addition.

\paragraph{Layout Design of the GLFA and VBFE} The proposed FoVB involves two well-designed modules, i.e., GLFA and VBFE. The GLFA module exploits forgery-aware convolutions to extract intra-modal forgery features, while the VBFE module leverages extracted features with variational Bayes to estimate audio-visual forgery traces in latent spaces. Both modules play crucial roles in enhancing forgery representation in continuous space. Specifically, in the shallow layers of the backbones, the GLFA modules are deployed to extract discriminative intra-modal forgery features, while in the deeper layers, the VBFE module is utilized to estimate audio-visual variables.

\subsection{Datasets}
\textbf{FakeAVCeleb \cite{khalid_fakeavceleb_2021}.} The FakeAVCeleb dataset contains $500$ real videos collected from different individuals of various ethnic groups, along with $19,500$ deepfake videos generated using various forgery algorithms, including FaceSwap \cite{marek_faceswap_2020} and FSGAN \cite{nirkin_fsgan_2019}, SV2TTS \cite{jia_transfer_2018}, and Wav2Lip \cite{prajwal_lip_2020}. Among these, SV2TTS is designed for audio forgery, while the others manipulate the visual contents. As a result, it contains five distinct categories of forgery samples as follows:
\begin{itemize}
    \item RVFA: Real Visuals - Fake Audio (SV2TTS)
    \item FVRA-WL: Fake Visuals - Real Audio (Wav2Lip)
    \item FVFA-WL: Fake Visuals - Fake Audio (SV2TTS + Wav2Lip)
    \item FVFA-FS: Fake Visuals - Fake Audio (SV2TTS + FaceSwap)
    \item FVFA-GAN: Fake Visuals - Fake Audio (SV2TTS + FSGAN)
\end{itemize}

\textbf{KoDF \cite{kwon_kodf_2021}.} This dataset collects massive videos of Korean speech. It comprises $62,166$ unique 90-second real video clips and $175,776$ deepfake clips, each lasting at least 15 seconds. The deepfake clips are generated using various algorithms, including FaceSwap, FSGAN, DeepFaceLab \cite{liu_deepfacelab_2023}, FOMM \cite{siarohin_first_2019}, ATFHP \cite{yi_audio-driven_2020}, and Wav2Lip. Note that we follow the evaluation setting of AVAD \cite{feng_self-supervised_2023} and use a subset of this dataset to evaluate the cross-dataset generalizability of our detector.

\textbf{DeAVMiT \cite{yang_avoid-df_2023}.} The DefakeAVMiT dataset is derived from VidTIMIT \cite{sanderson2002vidtimit}, which contains real videos, and then applies various forgery techniques to generate deepfake videos with subtle anomalies. Overall, it contains $540$ real videos and $6,480$ deepfake videos.

\textbf{DFDC \cite{dolhansky_deepfake_2020}.} The Deepfake Detection Challenge (DFDC) dataset consists of over $100,000$ video clips forged by various algorithms, e.g., FaceSwap, FSGAN, StyleGAN \cite{karras_style-based_2019}, TTS Skins \cite{polyak2019tts}, etc. Additionally, all videos are post-processed with various distortions, e.g., color saturation, blurring, compression, etc. Following the protocol in \cite{oorloff_avff_2024}, we utilize a subset of DFDC comprising $3,215$ audio-visual samples to assess the generalizability of the detector.

\textbf{LAV-DF \cite{cai_glitch_2023}.} The LAV-DF dataset comprises $36,431$ authentic videos from $153$ subjects and $99,873$ synthetic segments generated using various forgery techniques, similar to FakeAVCeleb. In audio forgery, word-level transcript manipulation is applied before SV2TTS to diversify the audio-visual inconsistencies.

\textbf{IDForge \cite{xu_identity-driven_2024}.} The IDForge dataset meticulously exploits a Large Language Model (LLM) to generate the false transcript at the sentence level, which exhibits fewer inconsistencies in the context compared with LAV-DF. Specifically, this dataset contains $214,428$ real videos collected from the Internet and generates challenging forgery samples via diverse combinations among visual, audio, and transcript modalities. Additionally, it provides a reference set containing extensive identity information to facilitate the detection of forgeries of specific identities, e.g., celebrities and political figures.

\subsection{Evaluation and Discussion}
Following the evaluation protocols in \cite{feng_self-supervised_2023, oorloff_avff_2024}, we evaluate the performance of our FoVB against existing counterparts on multiple dimensions, i.e., intra-dataset performance, cross-manipulation generalizability, cross-dataset generalizability, and robustness performance, on the FakeAVCeleb, KoDF, DeAVMiT, and DFDC datasets. For evaluation metrics, we report the accuracy (ACC), average precision (AP), and area under the receiver operating characteristic curve (AUC).

\paragraph{Intra-dataset Performance} Tab. \ref{tab: intra-dataset} demonstrates the comparative results regarding ACC and AUC metrics, which reveal the detectors' capability of capturing audio-visual artifacts. Specifically, it is observed as follows: (1) Overall, audio-visual methods, which leverage audio-visual anomalies to detect subtle forgeries, outperform uni-modal approaches that rely on artifacts from a single modality. (2) Compared to existing fine-tuned detectors, e.g., AVoiD-DF \cite{yang_avoid-df_2023} and MCL \cite{liu_mcl_2023}, which are directly fine-tuned with forgery-oriented datasets, our FoVB benefits from forgery-aware adaptation via variational Bayesian estimation and outperforms them with clear margins ($12.2\%$ in ACC, $10.2\%$ in AUC). (3) Moreover, compared with self-supervised detectors, e.g., PVASS-MDD \cite{yu_pvass-mdd_2023} and AVFF \cite{oorloff_avff_2024}, that require extra audio-visual pre-training on massive samples, our FoVB requires less reliance on large-scale pre-training datasets and efficiently estimates continuous latent spaces with variational Bayes efficiently, resulting in comparable detection performance.

\begin{table}[htb]
\centering
\caption{\textbf{Intra-Dataset Performance.} We report our method against recent methods regarding the ACC (\%)/AUC (\%) metrics. The best result is in bold, and the second best is underlined. $\dagger$ denotes the uni-modal detector and \textcolor{blue}{AVP} means audio-visual pre-training.}
\label{tab: intra-dataset}
\begin{tabular}{rccc}
\hline
\toprule[0.5pt]
Method & \textcolor{blue}{AVP} & FakeAVCeleb & KoDF \\ \hline
LipForensics$^{\dagger}$ \cite{haliassos_lips_2021} &  & 80.1 / 82.4 & 93.2 / 93.7 \\
FTCN$^{\dagger}$ \cite{zheng_exploring_2021}&  & 64.9 / 84.0 & 93.2 / 93.0 \\
RealForensics$^{\dagger}$ \cite{haliassos_leveraging_2022} & \textcolor{blue}{$\checkmark$} & 90.1 / 92.3 & -/- \\
MDS \cite{chugh_not_2020}&  & 82.8 / 86.5 & 95.7 / 95.2 \\
AV-DFD \cite{zhou_joint_2021}&  & 82.5 / 83.3 & 93.0 / 93.6 \\
AVoiD-DF \cite{yang_avoid-df_2023}&  & 83.7 / 89.2 & -/- \\
MCL \cite{liu_mcl_2023}&  & 86.0 / 89.6 & \underline{97.8} / \underline{98.1} \\
PVASS-MDD \cite{yu_pvass-mdd_2023}& \textcolor{blue}{$\checkmark$} & 95.7 / 97.3 & -/- \\
AVFF \cite{oorloff_avff_2024}& \textcolor{blue}{$\checkmark$} & \textbf{98.6} / \underline{99.1} & -/- \\ 
FRADE \cite{nie2024frade} & & 98.6 / 99.8 & 99.1 / 99.8 \\
PIA \cite{datta2025pia} & & 98.7 / 99.8 & -/- \\ \hline
FoVB (Ours) &  & \underline{98.5} / \textbf{99.7} & \textbf{99.1} / \textbf{99.8} \\ 
\bottomrule[0.5pt]
\hline
\end{tabular}
\end{table}

\paragraph{Cross-Manipulation Generalization} In real-world applications, the generalizability of detectors to unseen manipulation techniques is essential to discriminate evolving audio-visual artifacts. Therefore, following the previous \cite{feng_self-supervised_2023, oorloff_avff_2024}, we partition FakeAVCeleb into five categories: RVFA, FVRA-WL, FVFA-FS, FVFA-GAN, and FVFA-WL. And detectors are evaluated on one category while being trained on the remaining ones. As illustrated in Tab. \ref{tab: cross-manipulation}, it is observed as follows: (1) Compared with visual-only detectors, the audio-visual detectors demonstrate superior generalizability in identifying inconsistencies caused by either visual or audio forgeries. (2) Benefiting from the factorized variables in the VBFE module and the introduced orthogonality constraint, our method could effectively identify contributions of intra-modal (audio or visual) artifacts and cross-modal inconsistencies, which facilitates our detector in capturing more intrinsic audio-visual forgery artifacts, thus outperforms the existing counterparts with the clear margins, especially on RVFA ($+5.5\%$ in AP and $+4.8\%$ in AUC).

\begin{table*}[htb]
\centering
\caption{\textbf{Cross-Manipulation Generalization on FakeAVCeleb.} We evaluate the cross-manipulation generalizability of the detectors regarding AP (\%)/AUC (\%) by leaving out one category for testing while training on the remaining. Here, we consider the five categories in FakeAVCeleb: (1) \textbf{RVFA}: RealVisual-FakeAudio (SV2TTS), (2) \textbf{FVRA-WL}: FakeVisual-RealAudio (Wav2Lip), (3) \textbf{FVFA-FS}: FakeVisual-FakeAudio (FaceSwap + Wav2Lip + SV2TTS), (4) \textbf{FVFA-GAN}: FakeVisual-FakeAudio (FSGAN + Wav2Lip + SV2TTS), and (5) \textbf{FVFA-WL}: FakeVisual-FakeAudio (Wav2Lip + SV2TTS). AVG-FV denotes the average performance on the above categories containing fake visuals. \textcolor{blue}{AVP} and Modality denote whether the detector involves audio-visual pre-training or relies on audio-visual learning.}
\label{tab: cross-manipulation}
\begin{tabular}{rcccccccccccccc}
\hline
\toprule[0.5pt]
Method & Modality & \textcolor{blue}{AVP} & RVFA & FVRA-WL & FVFA-FS & FVFA-GAN & FVFA-WL & AVG-FV \\ \hline
Xception \cite{rossler_faceforensics_2019}& \textcolor{orange}{$\mathcal{V}$} &  & -/- & 88.2 / 88.3 & 92.3 / 93.5 & 67.6 / 68.5 & 91.0 / 91.0 & 84.8 / 85.3 \\
LipForensics \cite{haliassos_lips_2021}& \textcolor{orange}{$\mathcal{V}$} &  & -/- & 97.8 / 97.7 & 99.9 / 99.9 & 61.5 / 68.1 & 98.6 / 98.7 & 89.4 / 91.1 \\
FTCN \cite{zheng_exploring_2021}& \textcolor{orange}{$\mathcal{V}$} &  & -/- & 96.2 / 97.4 & 100. / 100. & 77.4 / 78.3 & 95.6 / 96.5 & 92.3 / 93.1 \\
RealForensics \cite{haliassos_leveraging_2022}& \textcolor{orange}{$\mathcal{V}$} & \textcolor{blue}{$\checkmark$} & -/- & 88.8 / 93.0 & 99.3 / 99.1 & \underline{99.8} / \underline{99.8} & 93.4 / 96.7 & 95.3 / 97.1 \\
AV-DFD \cite{zhou_joint_2021}& \textcolor{green}{$\mathcal{AV}$} &  & 74.9 / 73.3 & 97.0 / 97.4 & \underline{99.6} / \underline{99.7} & 58.4 / 55.4 & \textbf{100.} / \textbf{100.} & 88.8 / 88.1 \\
AVAD (LRS2) \cite{feng_self-supervised_2023}& \textcolor{green}{$\mathcal{AV}$} & \textcolor{blue}{$\checkmark$} & 62.4 / 71.6 & 93.6 / 93.7 & 95.3 / 95.8 & 94.1 / 94.3 & 93.8 / 94.1 & 94.2 / 94.5 \\
AVAD (LRS3) \cite{feng_self-supervised_2023}& \textcolor{green}{$\mathcal{AV}$} & \textcolor{blue}{$\checkmark$} & 70.7 / 80.5 & 91.1 / 93.0 & 91.0 / 92.3 & 91.6 / 92.7 & 91.4 / 93.1 & 91.3 / 92.8 \\
AVFF \cite{oorloff_avff_2024}& \textcolor{green}{$\mathcal{AV}$} & \textcolor{blue}{$\checkmark$} & \underline{93.3} / \underline{92.4} & 94.8 / 98.2 & \textbf{100.} / \textbf{100.} & \textbf{99.9} / \textbf{100.} & 99.4 / \underline{99.8} & \underline{98.5} / \underline{99.5} \\ 
ICSAV \cite{anshul2025intra} & \textcolor{green}{$\mathcal{AV}$} & \textcolor{blue}{$\checkmark$} & 94.0 / 95.4 & -/- & 99.3 / 96.6 & 99.3 / 95.2 & 99.0 / 90.9 & 99.2 / 94.2 \\ \hline
FoVB (Ours) & \textcolor{green}{$\mathcal{AV}$} & & \textbf{99.8} / \textbf{97.2} & \textbf{99.7} / 92.6 & 99.2 / 99.6 & 99.4 / 99.5 & \textbf{100.} / \underline{99.8} & \textbf{99.6} / \textbf{99.6} \\ 
\bottomrule[0.5pt]
\hline
\end{tabular}
\end{table*}

\paragraph{Cross-Dataset Generalization} It is also essential for practical detectors to generalize across different audio-visual distributions. Following recent methods \cite{yang_avoid-df_2023, yu_pvass-mdd_2023, oorloff_avff_2024}, we evaluate the generalizability of our FoVB on various datasets and report comparative results in Tab. \ref{tab: cross-dataset}. Specifically, audio-visual detectors achieve better generalization performance than visual-only detectors, which indicates cross-modal artifacts, i.e. inconsistencies, are intrinsic and universal across different audio-visual distributions. Typically, different distributions exhibit different audio-visual correlations, which could hinder the generalization capabilities of existing detectors. In contrast, audio-visual variables in our FoVB fully consider the dynamic changes of audio-visual correlations during estimation. By incorporating generalized prior knowledge and leveraging variational Bayes, our method achieves superior generalization performance ($+7.0\%$ in DeAVMiT and $+0.5\%$ in DFDC).

\begin{table}[htb]
\centering
\caption{\textbf{Cross-Dataset Generalization on KoDF, DFDC, and DeAVMiT.} We evaluate the generalizability of our detector trained on FakeAVCeleb regarding the AUC (\%) metric. The best result is in bold, and the second best is underlined. $\dagger$ denotes the uni-modal detector and \textcolor{blue}{AVP} means audio-visual pre-training.}
\label{tab: cross-dataset}
\begin{tabular}{rcccc}
\hline
\toprule[0.5pt]
Method & \textcolor{blue}{AVP} & KoDF & DFDC & DeAVMiT \\ \hline
LipForensics$^{\dagger}$ \cite{haliassos_lips_2021} &  & 86.6 & 53.1 & 52.5 \\
FTCN$^{\dagger}$ \cite{zheng_exploring_2021} &  & 68.1 & - & - \\
MDS \cite{chugh_not_2020} &  & - & 73.1 & 75.2 \\
AV-DFD \cite{zhou_joint_2021}&  & - & 76.7 & 77.8 \\
AVoiD-DF \cite{yang_avoid-df_2023} &  & - & 80.7 & 83.2 \\
PVASS-MDD \cite{yu_pvass-mdd_2023} & \textcolor{blue}{$\checkmark$} & - & 84.8 & \underline{87.5} \\
AVAD \cite{feng_self-supervised_2023} & \textcolor{blue}{$\checkmark$} & 86.9 & 81.4 & 83.7 \\
AVFF \cite{oorloff_avff_2024} & \textcolor{blue}{$\checkmark$} & \textbf{95.5} & \underline{86.2} & - \\ 
FRADE \cite{nie2024frade} & & 92.4 & 83.8 & 89.3 \\
ICSAV \cite{anshul2025intra} &\textcolor{blue}{$\checkmark$} & 99.2 & 85.2 & - \\
\hline
FoVB (Ours) &  & \underline{94.3} & \textbf{86.7} & \textbf{90.7} \\ 
\bottomrule[0.5pt]
\hline
\end{tabular}
\end{table}

\begin{figure}[htbp]
    \centering
    \includegraphics[width=\linewidth]{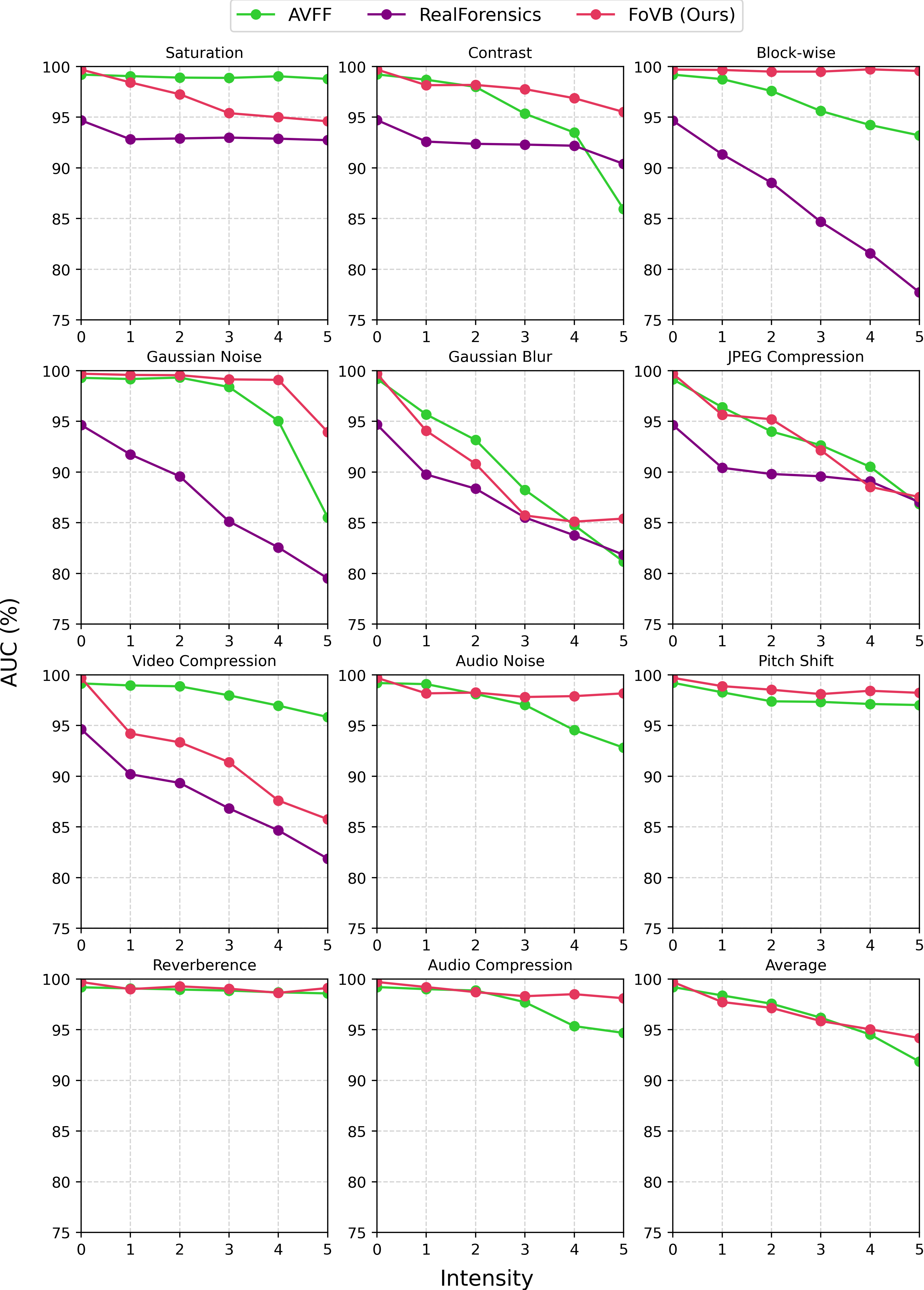}
    \caption{\textbf{Robustness to Unseen Perturbations.} Following the setting of AVFF, we report AUC scores (\%) for various perturbations with different intensities evaluated on the test set of FakeAVCeleb. Specifically, the first seven perturbations attack visual contents, while the remaining ones are audio perturbations.}
    \label{fig: robustness}
\end{figure}

\paragraph{Robustness to Unseen Perturbations} In real-world scenarios, deepfake videos typically suffer severe post-processing operations, i.e., perturbations, when uploaded to social media platforms or stored on various mediums, which distort and undermine their artifacts. Hence, the detectors are expected to be robust to unseen perturbations. Following previous arts \cite{haliassos_lips_2021, haliassos_leveraging_2022, feng_self-supervised_2023, oorloff_avff_2024}, we evaluate our FoVB on diverse audio-visual perturbations with multiple intensities. The results are illustrated in Fig.\ref{fig: robustness} and analyzed as follows: (1) Overall, the audio-visual methods, i.e., AVFF and our FoVB, achieve better robustness performance with significant improvements, suggesting that cross-modal artifacts indeed facilitate the learning of robust forgery representation. (2) Benefiting from diverse latent Gaussian variables and the variable $f_{K}$ introduced in ELBO optimization, our FoVB could effectively estimate the accurate latent variable $c$ even if modality perturbations exist, and outperform the counterparts on most types of perturbations. (3) However, our method is hindered from extracting discriminative artifacts by video compression perturbation, which primarily damages visual temporal information and thus interferes with the estimation of cross-modal inconsistencies. In short, the proposed FoVB exhibits promising robustness performance for the detection of distorted audio-visual deepfake content, enabling it to be deployed in real-world scenarios, e.g., live streaming and social media sharing.

\paragraph{Additional Evaluations on Recent Benchmarks}
To further demonstrate the superiority of the proposed FoVB framework against evolving forgery techniques, we conduct several experiments on recently published datasets, e.g., LAV-DF and IDForge, regarding both intra-dataset and cross-dataset evaluation. Specifically, some forged videos in LAV-DF contain both real and fake segments, which could be detected by temporal detection methods and are out of our scope. Therefore, we would extract the corresponding fake segments as fake samples. Meanwhile, some forged videos in IDForge are identified as problematic videos, with facial regions either distorted or difficult to detect. These videos will be excluded during our evaluation. The results are reported in Tab. \ref{tab: supp_evaluation}. It can be observed that the proposed FoVB achieves satisfactory detection performance on the intra-dataset evaluations of LAV-DF and IDForge. This indicates that our method could effectively learn discriminative audio-visual forgery representation regardless of diverse forgery types. Furthermore, our FoVB exhibits better generalizability to LAV-DF on cross-dataset evaluation than IDForge, which contains more complicated forgeries spanning audio, visual, and text modalities. This highlights the importance of incorporating more comprehensive modality information, beyond just audio-visual modalities, into the design of an ideal forgery detector.

\begin{table}
\centering
\caption{\textbf{Additional Evaluations on Recent Benchmarks.} In cross-dataset evaluation, our FoVB trained on FakeAVCeleb is evaluated on the following datasets. To the best of our knowledge, none of the previous arts reports the cross-dataset performance results on the above benchmarks.}
\label{tab: supp_evaluation}
\begin{tabular}{lcccc}
\hline
\toprule[0.5pt]
\multirow{2}{*}{Dataset} & \multicolumn{2}{c}{Intra-dataset} & \multicolumn{2}{c}{Cross-dataset} \\
 & AP(\%) & AUC(\%) & AP(\%) & AUC(\%) \\ \hline
LAV-DF  & 97.6 & 98.1 & 90.7 & 89.2\\
IDForge  & 90.3 & 95.6 & 83.4 & 78.3 \\
\bottomrule[0.5pt]
\hline
\end{tabular}
\end{table}

\begin{figure}
    \centering
    \includegraphics[width=\linewidth]{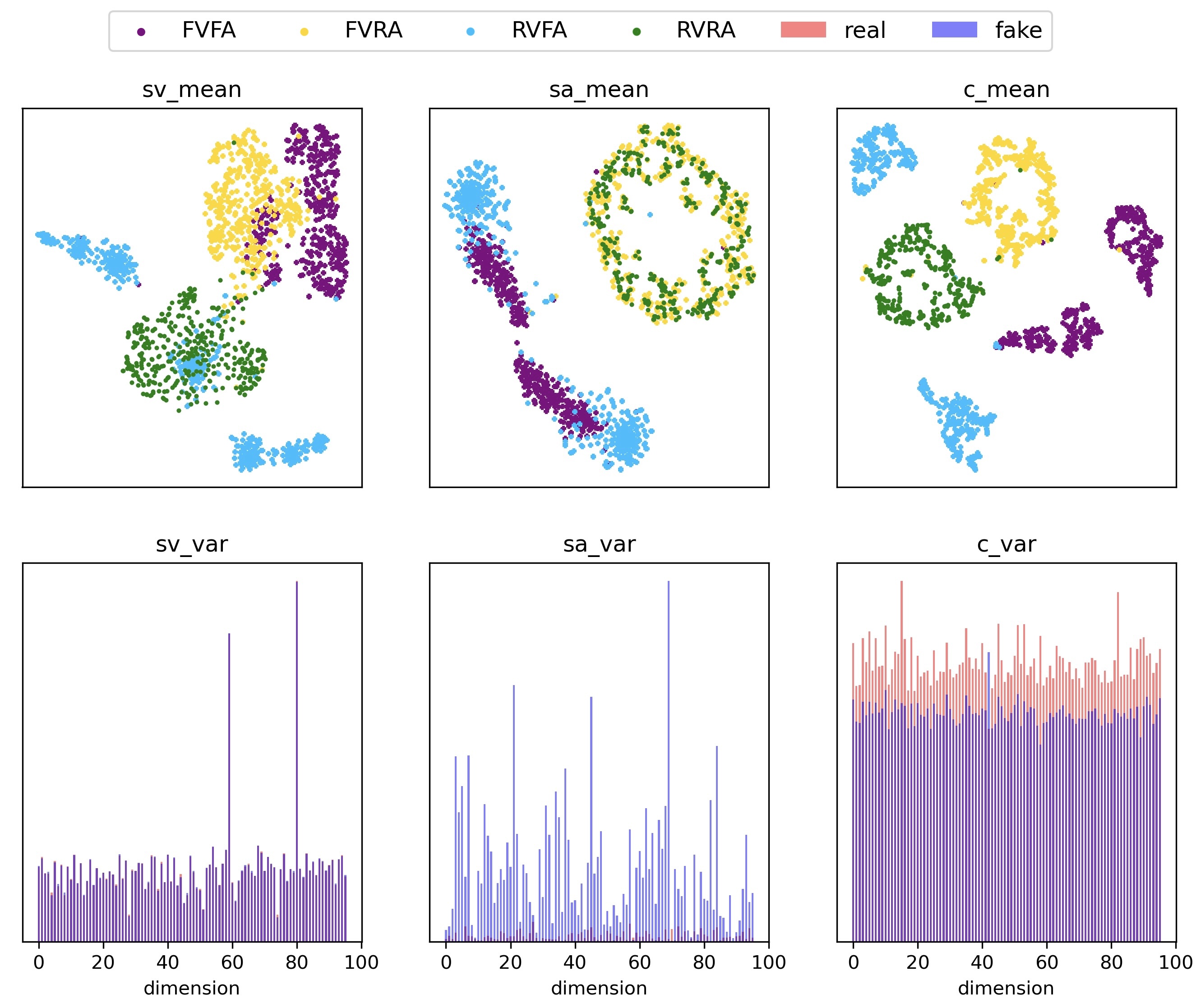}
    \caption{\textbf{Analysis of the Factorized Variables.} In the first row, based on forgery categories, we visualize the mean values of audio-visual samples via t-SNE projection while we present average variance values in each feature dimension.}
    \label{fig: variable_vis}
\end{figure}

\paragraph{Analysis on the Factorized Latent Variables.}
Driven by our method's promising generalizability and robustness, we delve into the factorized variables estimated by variational Bayes in VBFE. Specifically, we visualize the t-SNE plots of the mean values in these Gaussian variables and statistic the corresponding variance values. It is observed in Fig.\ref{fig: variable_vis} as follows: (1) The mean values of modality-specific variables $s_{a}, s_{v}$ are clustered based on their modality authenticity with minimal entanglement, while the correlation-specific variable $c$ is clustered based on their combination categories, which implies different combinations, i.e., RVFA, FVRA, and FVFA, could produce exclusive audio-visual inconsistencies. Hence, desirable generalizability comes from identifying natural audio-visual correlations effectively, rather than focusing on any specific inconsistencies. (2) Compared with visual artifacts, audio and cross-modal artifacts could cause clear discrepancies in the variance values of the corresponding variables. It indicates that cross-modal artifacts actually exhibit lower diversity and are more suitable to be regarded as generalized artifacts in audio-visual deepfake detection.

\subsection{Ablation Studies}

\begin{table}
\centering
\caption{\textbf{Evaluation on Ablations.} We evaluate diverse variants of our FoVB within intra-dataset (FakeAVCeleb), cross-manipulation (RVFA), and cross-dataset (KoDF) settings regarding the metric AUC (\%). The best result is in bold, and the second best is underlined.}
\label{tab: ablation}
\begin{tabular}{lccc}
\hline
\toprule[0.5pt]
Method &FakeAVCeleb & RVFA & KoDF \\ \hline
w/o GLFA & 99.8 & 96.8 & 93.5 \\
w/o VBFE & 99.4 & 92.1 & 91.6 \\ \hline
Only single local & 98.4 & 94.2 & 91.3 \\
Only global & 98.1 & 95.7 & 92.5 \\
Ours w/o factorization & \underline{99.6} & 93.8 & 92.4 \\
Ours w/o dynamic prior $f_{K}$ & 99.4 & 92.0 & 91.7 \\
Ours w/o $\mathcal{L}_{\mathrm{orth}}$ & \textbf{99.7} & 95.1 & 93.6 \\ \hline
Ours w/ fewer samples & 99.4 & 95.6 & 92.1 \\
Ours w/ large backbone & \textbf{99.7} & \textbf{97.8} & \textbf{95.9} \\ \hline
FoVB (Ours) & \textbf{99.7}  & \underline{97.2} & \underline{94.3} \\ 
\bottomrule[0.5pt]
\hline
\end{tabular}
\end{table}

\paragraph{GLFA and VBFE Collaboration} We replace GLFA and VBFE modules with vanilla convolutions and simple cross-attention layers, respectively, to explore their contributions regarding generalizability and robustness. The results are shown in Tab. \ref{tab: ablation}. Specifically, GLFA mainly contributes to performance improvement in intra-dataset evaluations, where discriminative forgery clues are required. In contrast, VBFE focuses on capturing more generalized forgery representations via variational Bayesian estimation. Moreover, compared with vanilla convolutions, forgery-aware convolutions in GLFA are more effective at extracting higher-quality forgery features, facilitating the estimation of audio-visual latent variables with variational Bayes in VBFE.

\paragraph{Diverse Forgery-aware Convolutions} GLFA exploits difference and high-frequency convolutions to extract intra-modal forgery features from local and global perspectives. Compared with vanilla convolution, they focus on mining forgery information in both local textural and high-frequency regions while minimally being affected with semantic information, thus improving the generalizability of our method. To verify the necessity of this design, we ablate involved convolutions and report the results in Tab. \ref{tab: ablation}. It can be observed that performance degradation occurs when either difference or high-frequency convolutions are discarded, which indicates that both local and global views could reveal distinct features of forgery traces. Moreover, among the different types of convolutions, the high-frequency convolution contributes more significantly to improving the generalizability of our method.

\begin{figure}
    \centering
    \includegraphics[width=0.8\linewidth]{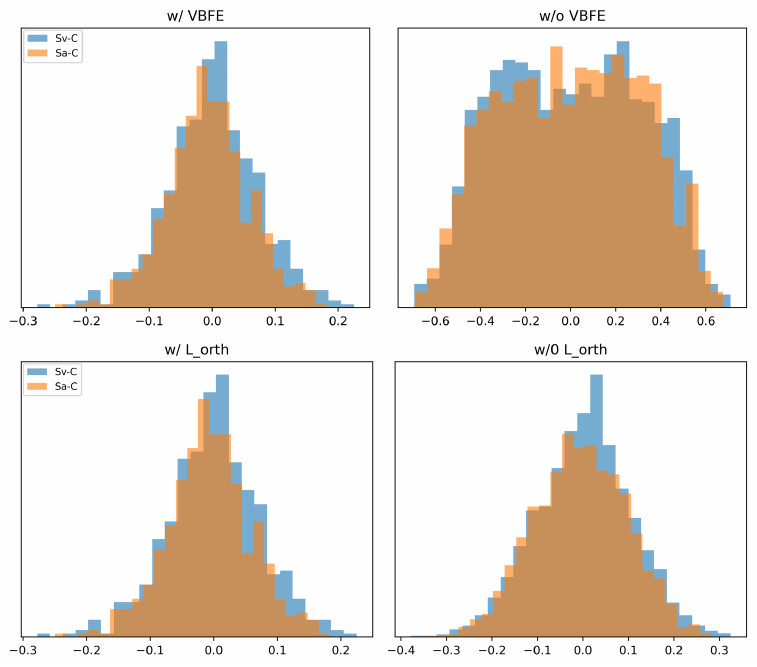}
    \caption{\textbf{Latent Similarity Comparison.} We collect audio-visual samples from FakeAVCeleb and compare the similarities between modality-specific latent variables and correlation-specific ones.}
    \label{fig: latent_sim}
\end{figure}

\paragraph{Audio-Visual Variable Factorization} As the core designs of the proposed VBFE module in our FoVB, both variable factorization and orthogonality constraint ($\mathcal{L}_{\mathrm{orth}}$) are exploited to estimate modality-specific and correlation-specific variables with less entanglement in latent spaces, which are crucial for generalization performance. To assess the contributions of these designs, we ablate all involved designs and report corresponding results in Tab. \ref{tab: ablation}. Compared with using only the $\mathcal{L}_{\mathrm{orth}}$, the variable factorization could achieve a promising performance improvement ($+3\%$ in AUC). It indicates that factorization-aware Bayesian estimation could facilitate the disentanglement of audio-visual artifacts, as opposed to $\mathcal{L}_{\mathrm{orth}}$ after variable estimation. Notably, when only $\mathcal{L}_{\mathrm{orth}}$ is applied, our FoVB projects the audio-visual estimated variable into three fine-grained sub-variables. In addition, we measure the latent similarities among $c$ and $s_{a}, s_{v}$ to intuitively validate the effectiveness of VBFE and $\mathcal{L}_{\mathrm{orth}}$. As shown in Fig.\ref{fig: latent_sim}, without the Bayesian estimation of audio-visual correlations in VBFE, the involved three latent variables are similar to each other, indicating that both audio and visual latent representations tend to capture the correlation-specific features rather than their own modality-specific ones. While $\mathcal{L}_{\mathrm{orth}}$ could help to learn the intra-modal and cross-modal forgery representations with less entanglement. We also explore the impacts of $\mathcal{L}_{\mathrm{orth}}$ in $\mathcal{L}$ with different weight values, as shown in Tab. \ref{tab: alpha_ablation}. It turns out that larger $\alpha$ values undermine the generalizability of our FoVB, that is, overestimating the disentanglement of different forgery traces hinders the trained detectors from capturing generalized forgery features.

\begin{table}[]
\centering
\caption{The $\alpha$ values in $\mathcal{L}$. We evaluate different values of $\alpha$ within intra-dataset (FakeAVCeleb), cross-manipulation (RVFA), and cross-dataset (KoDF) settings regarding the metric AUC (\%). The best result is in bold, and the second best is underlined.}
\label{tab: alpha_ablation}
\begin{tabular}{lccc}
\hline
\toprule[0.5pt]
$\alpha$ & FakeAVCeleb & RVFA & KoDF \\ \hline
0.1 & \underline{99.7} & \textbf{97.2} & \underline{94.3} \\
0.2 & 99.5 & 96.3 & \textbf{94.7} \\
0.4 & \textbf{99.8} & 95.2 & 93.2 \\
0.6 & \underline{99.7} & \underline{96.4} & 91.8 \\
0.8 & 99.4 & 95.8 & 90.5 \\ 
\bottomrule[0.5pt]
\hline
\end{tabular}
\end{table}

\paragraph{Scalability and Training Efficiency} 
Our FoVB adapts the vanilla ViT as the default backbone for relatively fair comparisons. Meanwhile, we exploit extremely limited samples, i.e., $2,000$ samples, to train the proposed framework. As shown in Tab. \ref{tab: ablation}, even with this limited number of samples, our FoVB achieves detection performance comparable to, or even better than, state-of-the-art methods such as AVFF and PAVSS-MDD, which rely on massive audio-visual samples (at least $97,000$ samples on LRS2) to estimate audio-visual correlations. Furthermore, we explore more advanced backbones with larger parameter volumes, e.g. ViT-large and SwinT-large \cite{liu_swin_2021}. The results in Tab. \ref{tab: ablation} indicate that our FoVB has the potential to learn more generalizable forgery representations and achieve better generalization performance when coupled with evolving backbones.

\begin{table}[ht]
    \centering
    \caption{\textbf{Deployment Location of VBFE.} 3-$rd$ denotes the VBFE exploits extracted audio-visual forgery features by the third transformer block of the ViT-base backbone, i.e., the shallow layer of the given backbone. Similarly, 6-$th$ and 9-$th$ represent VBFE is deployed into the intermediate and deep layers of the given backbone.}
    \label{tab: location_ablation}
    \begin{tabular}{ccccc}
    \hline
    \toprule[0.5pt]
        Loc. & FakeAVCeleb & RVFA & KoDF & AVG-R \\ \hline
        3-$rd$ & \textbf{99.8} & 92.6 & 89.5 & 87.2 \\
        6-$th$ & \underline{99.7} & \textbf{97.2} & \textbf{94.3} & \underline{94.1} \\
        9-$th$ & \underline{99.7} & \underline{96.3} & \underline{93.7} & \textbf{94.8} \\ 
    \bottomrule[0.5pt]
    \hline
    \end{tabular}
\end{table}

\begin{table}[ht]
    \centering
    \caption{\textbf{Latent Variable Adaptation Choice.} X-attn, Concat, and Addition denote the sampled variables are fused with raw audio-visual features in the backbone via cross-attention, concatenation, and element-wise addition, respectively.}
    \label{tab: injection_ablation}
    \begin{tabular}{lcccc}
    \hline
    \toprule[0.5pt]
        Injection & FakeAVCeleb & RVFA & KoDF & AVG-R \\ \hline
        X-attn & \textbf{99.7} & \textbf{97.3} & \underline{94.1} & 93.7 \\
        Concat & \underline{99.6} & 96.5 & 93.2 & \textbf{94.8} \\
        Addition & \textbf{99.7} & \underline{97.2} & \textbf{94.3} & \underline{94.1}  \\
    \bottomrule[0.5pt]
    \hline
    \end{tabular}
\end{table}

\begin{figure}
    \centering
    \includegraphics[width=0.8\linewidth]{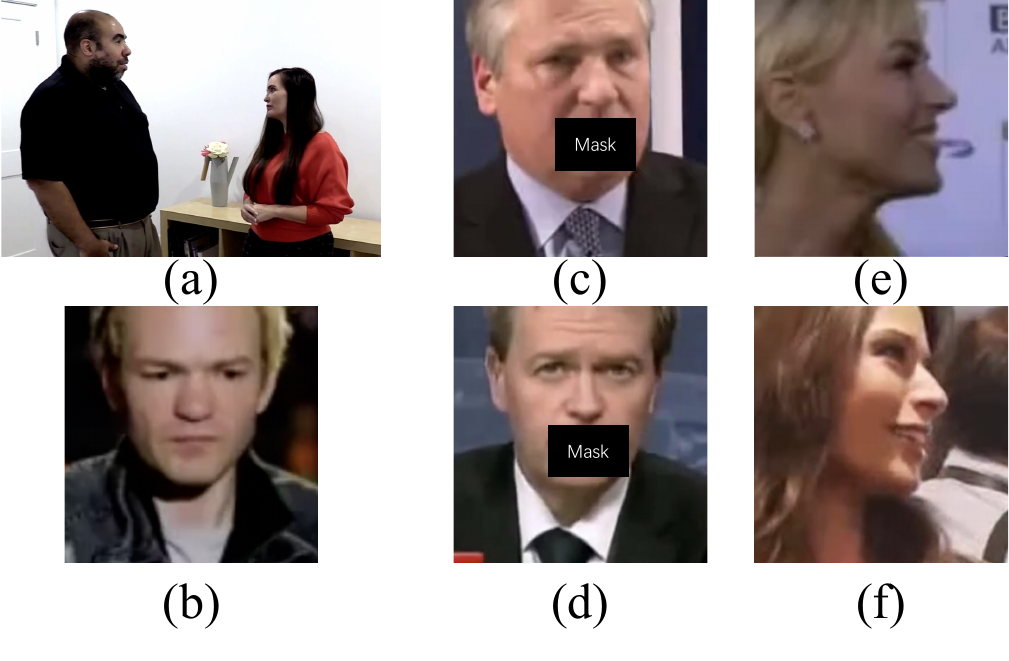}
    \caption{\textbf{Challenging Detection Examples.} (a) Multiple speakers with mixed audio signals. (b) Stationary speaker. (c)-(d) Partially facial occlusion. (e)-(f) The extreme head posture.}
    \label{fig: failture}
\end{figure}

\paragraph{Deployment Location of VBFE} According to the design principle of VBFE, the audio-visual features extracted at different layers would affect the effectiveness of variational Bayesian estimation. To determine the optimal deployment of VBFE, we perform an ablation study on different layer locations (i.e., the 3rd, 6th, and 9th layers, representing shallow, intermediate, and deep layers, respectively). The results are illustrated in Tab. \ref{tab: location_ablation} and observed as follows. VBFEs deployed at different locations consistently achieve promising intra-dataset performance on FakeAVCeleb, though their generalizability varies. Specifically, using features extracted from deeper layers brings significant improvement in the cross-dataset evaluation, and the best cross-manipulation performance is obtained when the VBFE is deployed at the intermediate location.

\paragraph{Latent Variable Adaptation Choice} Given the sampled audio-visual variables ${sc}_{a}$ and ${sv}_{v}$, we explore adapting ${sc}_{a}$ and ${sv}_{v}$ into pre-trained backbone in different manners, and report the results in Tab. \ref{tab: injection_ablation}. It is shown that all adaptation methods could achieve comparable detection performance across multiple evaluations. Therefore, due to its simplicity, we select element-wise addition as the default adaptation of audio-visual variables in our FoVB.

\subsection{Challenging Detection Examples}
Since the estimation of the correlation-specific variable $c$ with VBFE heavily relies on the audio-visual correspondence, i.e., the relationship between the audio track and lip region, learning cross-modal inconsistencies becomes challenging in scenarios where the direct audio-visual correspondence could not be established. Fig.\ref{fig: failture} illustrates the following four typical scenarios of detection failure: (1) Multiple speakers: when audio tracks from different speakers are mixed up, it is difficult to establish the audio-visual correspondence for any single speaker, as shown in Fig.\ref{fig: failture}(a). (2) Stationary speaker: the absence of clear lip movement prevents the detector from capturing meaningful forgery traces as shown in Fig.\ref{fig: failture}(b). (3) Partially facial occlusion: the occlusion of lip movements, which are crucial for establishing audio-visual correlation, leads to the detection failure as shown in Fig.\ref{fig: failture}(c) and (d). (4) The extreme head posture: the abnormal head postures, e.g., side face, make it difficult to capture the clue of lip movenments as shown in Fig.\ref{fig: failture}(e) and (f).

\section{Conclusion}
This paper proposes a novel framework, i.e., Forgery-aware Audio-Visual Adaptation with Variational Bayes (FoVB), for multi-modal deepfake detection, which adapts forgery-relevant knowledge of audio-visual correlations into the pre-trained backbones with variational Bayes. First, we exploit well-devised difference convolutions and the high-pass filter to extract intra-modal forgery features from local and global perspectives. Second, we formulate the learning of generalizable representation of audio-visual correlations as latent variable estimation with variational Bayes to delve into the rich semantics in the latent space. In particular, the latent variable is decomposed into modality-specific and correlation-specific variables with the orthogonal constraint to minimize the entanglement with each other while better learning the intra-modal and cross-modal forgery representations. Extensive experiments demonstrate the effectiveness of our designs and show that the performance of the proposed method surpasses other state-of-the-art competitors in terms of generalizability and robustness.

\appendices
\counterwithin{equation}{section}
\counterwithin{figure}{section}
\renewcommand{\theequation}{\thesection-\arabic{equation}}
\renewcommand{\thefigure}{\thesection-\arabic{figure}}

\section{GLFA Kernel Details}
\label{supp: kernel}
\begin{figure}[htbp]
    \centering  
    \includegraphics[width=\linewidth]{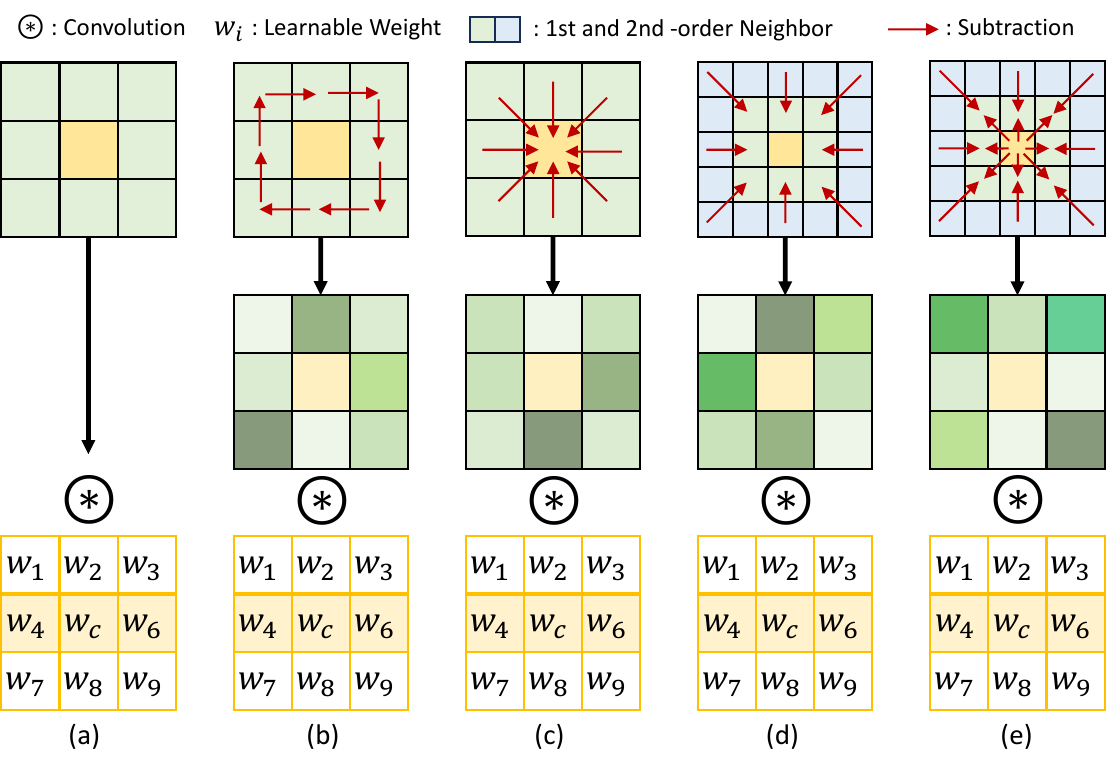}
    \caption{Illustration of the introduced Difference Convolutions in GLFA module. (a) Vanilla Convolution. (b) Angular Difference Convolution. (c) Central Difference Convolution. (d) Radial Difference Convolution. (e) Second-order Convolution.}
    \label{fig: GLFA_kernels}
\end{figure}

\paragraph{Difference Forgery-aware Convolutions} As shown in Fig.\ref{fig: GLFA_kernels}, we decompose difference convolutions into two-stage operations \cite{liuExtendedLocalBinary2012, liuFusingSortedRandom2015, suBIRDLearningBinary2019}. The audio (visual) features are processed with diverse subtraction operations and then convoluted with the learnable weights $w$.

\paragraph{High-pass Filter $M_{f}$} In the frequency domain, the low-frequency components of audio (visual) features concentrate around central regions. Hence, to discard the low-frequency components, the high-pass filter $M_{f}$ is initialized to zero, where the central area with radius $d_{f}$ is set as zero and performs element-wise multiplication with transformed features. Note that $d_{f}$ is set as one-quarter of the corresponding feature shapes.

\section{Variational Inference}
\label{supp: cond_vae}
Given the posterior approximation $q_{\phi}(z \vert X, y)$, we could derive the objective $p_{\theta}(y \vert X)$ as follows:

{\footnotesize
\begin{flalign}
    \begin{split}
        \mathrm{log}p_{\theta}(y \vert X) &= \left [\int_{z} q_{\phi}(z \vert X, y) dz \right] \mathrm{log}p_{\theta}(y \vert X)  \\
        &= \mathbb{E}_{q_{\phi}(z \vert X, y)} \mathrm{log} p_{\theta}(y \vert X) \\
        &= \mathbb{E}_{q_{\phi}(z \vert X, y)} \mathrm{log} \frac{p_{\theta}(z, X, y) \, p_{\theta}(y \vert X)}{p_{\theta}(z, X, y)} \\
        &= \mathbb{E}_{q_{\phi}(z \vert X, y)} \mathrm{log} \frac{p_{\theta}(y \vert X, z) \, p_{\theta}(z \vert X) \, p_{\theta}(X) \, p_{\theta}(y \vert X)}{p_{\theta}(z \vert X, y) \, p_{\theta}(X, y)} \\
        &= \mathbb{E}_{q_{\phi}(z \vert X, y)} \mathrm{log} \frac{q_{\phi}(z \vert X, y)}{p_{\theta}(z \vert X, y)} + \mathbb{E}_{q_{\phi}(z \vert X, y)} \mathrm{log} p_{\theta}(y \vert X, z) \\
        &-\mathbb{E}_{q_{\phi}(z \vert X, y)} \mathrm{log} \frac{q_{\phi}(z \vert X, y)}{p_{\theta}(z \vert X)} \\
        &= D_{KL}(q_{\phi}(z \vert X, y) \Vert p_{\theta}(z \vert X, y)) + \mathbb{E}_{q_{\phi}(z \vert X, y)} \mathrm{log} p_{\theta}(y \vert X, z) \\
        &- D_{KL}(q_{\phi}(z \vert X, y) \Vert p_{\theta}(z \vert X)), \\
    \end{split}
    \label{equ: elbo_1}
\end{flalign}
}

\noindent where $D_{KL}$ means the Kullback-Leibler (KL) divergence. Moreover, $D_{KL}(q_{\phi}(z \vert X, y) \vert p_{\theta}(z \vert X, y))$ in Eq. \ref{equ: elbo_1} is always greater or equal to zero. Therefore, the evidence lower bound (ELBO) of Eq. \ref{equ: elbo_1} could be derived as:

{\footnotesize
\begin{flalign}
    \begin{split}
        \mathrm{log}p_{\theta}(y \vert X) & \geq \mathrm{ELBO}(X, y, \theta, \phi) \\
        &= \mathbb{E}_{q_{\phi}(z \vert X, y)} \mathrm{log} p_{\theta}(y \vert X, z) - D_{KL}(q_{\phi}(z \vert X, y) \Vert p_{\theta}(z \vert X)), \\
    \end{split}
    \label{equ: elbo_2}
\end{flalign}
}

\noindent where the first term is to measure the reconstruction quality, and the second term is utilized to compute the KL divergence from the approximate posterior $q_{\phi}(z \vert X, y)$ to the prior $p_{\theta}(z \vert X)$.

\section{Multi-modal Variational Inference}
\label{supp. dynamic prior}

With the defined $f_{K}$, we can rewrite the low bound of $D_{KL}$ term in Eq. \ref{equ:low_bound_moe} as:

{\footnotesize
\begin{flalign}
    \begin{aligned}
        &\ \sum_{i=1}^{K} \pi_{i} D_{KL}(q_{\phi_{i}}(z \vert x_{i}, y) \Vert p_{\theta}(z \vert x_{i})) &\\
        &\ \leq \sum_{i=1}^{K} \pi_{i} D_{KL}(q_{\phi_{i}}(z \vert x_{i}, y) \Vert f_{K}) &\\
        &\ \leq \frac{1}{2} \sum_{i=1}^{K} \pi_{i} D_{KL}(q_{\phi_{i}}(z \vert x_{i}, y) \Vert f_{K}) + \frac{1}{2} \sum_{i=1}^{K} \pi_{i} D_{KL}(p_{\theta}(z \vert x_{i}) \Vert f_{K}), &\\
    \end{aligned}
    \label{equ: fk_KLmoe}
\end{flalign}
}
\noindent where the first inequality stands because compared with the uni-modal prior $p_{\theta}(z \vert x_{i})$, the $f_{K}$ describes the more accurate distribution of the shared variable $z$ using multiple posteriors and priors, which enlarges its KL discrepancy from $q_{\phi_{i}}(z \vert x_{i}, y)$. Moreover, compared with $q_{\phi_{i}}(z \vert x_{i}, y)$, the prior $p_{\theta}(z \vert x_{i})$ could not capture the mutli-modal characteristics of latent variable $z$, i.e., without the posterior information (label $y$). That leads to the larger KL discrepancy from $p_{\theta}(z \vert x_{i})$ to $f_{K}$, thus, the second inequality holds.

Given the two combinatorial distributions $P(x) = \sum^{m}_{i=1} \pi^{P}_{i} P_{i}(x)$ and $Q(x) = \sum_{i=1}^{n} \pi^{Q}_{i} Q_{i}(x)$, with $\sum^{m}_{i=1} \pi^{Q}_{i} = 1$ and $\sum^{n}_{i=1} \pi^{P}_{i} = 1$, the Jensen-Shannon divergence $D_{JS}$ is defined as follows:

{\footnotesize
\begin{gather}
    M(x) = \frac{P(x) + Q(x)}{2}, \\
    D_{JS}(P \Vert Q) = \frac{1}{2} \sum_{i} \pi^{P}_{i} D_{KL}(P_{i} \Vert M) + \frac{1}{2} \sum_{i} \pi^{Q}_{i} D_{KL}(Q_{i} \Vert M).
    \label{equ: mmjsd}
\end{gather}
}

\noindent Therefore, combining with the MoE-based $f_{K}$ and Eq. \ref{equ: mmjsd} , the lower bound of $\mathrm{ELBO}(X,y,\theta,\phi)$, i.e., $\widetilde{\mathrm{{ELBO}}}(X,y,\theta,\phi)$, could be derived as follows.

{\footnotesize
\begin{flalign}
    \begin{aligned}
        &\ \mathrm{ELBO}(X,y,\theta,\phi) &\\
        &\ \geq \mathbb{E}_{q_{\phi}(z \vert X, y)} \mathrm{log} \, p_{\theta}(y \vert X, z) - \sum_{i=1}^{K} \pi_{i} D_{KL}(q_{\phi_{i}}(z \vert x_{i},y) \Vert f_{K}) &\\
        &\ \geq \mathbb{E}_{q_{\phi}(z \vert X, y)} \mathrm{log} \, p_{\theta}(y \vert X, z)- \frac{1}{2} \sum_{i=1}^{K} \pi_{i} D_{KL}(q_{\phi_{i}}(z \vert x_{i},y) \Vert f_{K}) &\\
        &\ - \frac{1}{2} \sum_{i=1}^{K} \pi_{i} D_{KL}(p_{\theta}(z \vert x_{i}) \Vert f_{K}) & \\
        &\ = \mathbb{E}_{q_{\phi}(z \vert X, y)} \mathrm{log} \, p_{\theta}(y \vert X, z) - D_{JS}(q_{\phi}(z \vert X,y), p_{\theta}(z \vert X)) & \\
        &\ = \widetilde{\mathrm{{ELBO}}}(X,y,\theta,\phi).
    \end{aligned}
    \label{equ: supp_js_lowbound}
\end{flalign}
}

\section{Audio-Visual Forgery-aware ELBO}
To estimate individual contributions of intra-modal and cross-modal forgery artifacts in variational Bayes, we factorize the universal variable $z$ into two types of independent variables, i.e., the modality-specific $s = \{s_{a}, s_{v}\}$ and correlation-specific $c$. Therefore, the reconstruction term in $\widetilde{\mathrm{ELBO}}(X,y,\theta,\phi)$ could be rewritten as:

{\footnotesize
\begin{flalign}
    \begin{aligned}
        &\ \mathbb{E}_{q_{\phi}(z \vert X, y)} \mathrm{log} p_{\theta}(y \vert X, z) & \\
        &\ = \mathbb{E}_{q_{\phi}(s,c \vert X, y)} \mathrm{log} p_{\theta}(y \vert X, s,c) & \\
        &\ = \int q_{\phi_{c}}(c \vert X,y) q_{\phi_{s}}(s \vert X, y) \mathrm{log} p_{\theta}(y \vert X, s,c) dc & \\
        &\ = \int q_{\phi_{c}}(c \vert X,y) \sum_{o \in \{a, v\}} \int_{s_{o}} q_{\phi_{o}}(s_{o} \vert x_{o}, y) \mathrm{log}p_{\theta}(y \vert s_{o}, c) ds_{o}dc & \\
        &\ = \sum_{o \in \{a, v\}} \mathbb{E}_{q_{\phi_{c}}(c \vert X, y)} \left[ \mathbb{E}_{q_{\phi_{s_{o}}}(s_{o} \vert x_{o}, y)} [\mathrm{log}p_{\theta}(y \vert s_{o}, c)]  \right]. &
    \end{aligned}
    \label{equ: factorized_recon}
\end{flalign}
}

Similarly, we could rewrite the KL-divergence, which includes $c$ using the MoE-based dynamic prior and the JS-divergence for joint distributions:

{\footnotesize
\begin{flalign}
    \begin{aligned}
        &\ D_{KL}(q_{\phi}(z \vert X,y) \Vert p_{\theta}(z \vert X)) & \\
        &\ = D_{KL}(q_{\phi}(s,c \vert X,y) \Vert p_{\theta}(s,c \vert X)) & \\
        &\ = D_{KL}(q_{\phi_{s}}(s \vert X, y) q_{\phi_{c}}(c \vert X, y) \Vert p_{\theta_{s}}(s \vert X) p_{\theta_{c}}(c \vert X))  & \\
        &\ = D_{KL}(q_{\phi_{s}}(s \vert X,y) \Vert p_{\theta_{s}}(s \vert X)) + D_{KL}(q_{\phi_{c}}(c \vert X,y) \Vert p_{\theta_{c}}(c \vert X)) & \\
        &\ = \sum_{o \in \{a,v\}} D_{KL}(q_{\phi_{s_{o}}}(s_{o} \vert x_{o},y) \Vert p_{\theta}(s_{o} \vert x_{o})) + D_{JS}(q_{\phi}(c \vert X,y), p_{\theta}(c \vert X)), &
    \end{aligned}
    \label{equ: factorized_js}
\end{flalign}
}
where the latent variable $z$ is decomposed into the modality-specific $s_{a}, s_{v}$, and the correlation-specific $c$. The former extracts intra-modal artifacts in latent spaces, while the variable $c$ captures cross-modal artifacts in the latent space.

\bibliographystyle{IEEEtran}
\bibliography{purified_A2MiC_reference}

\end{document}